\documentclass[conference]{IEEEtran}
\IEEEoverridecommandlockouts
\usepackage{cite}
\usepackage{amsmath,amssymb,amsfonts}
\usepackage{algorithmic}
\usepackage{graphicx}
\usepackage{textcomp}
\usepackage{xcolor}
\usepackage{subfigure}
\usepackage{booktabs}
\usepackage{multirow}
\usepackage{threeparttable}
\usepackage{makecell}
\usepackage{balance}
\usepackage[ruled, linesnumbered, vlined]{algorithm2e}
\usepackage{algorithmic}

\def\BibTeX{{\rm B\kern-.05em{\sc i\kern-.025em b}\kern-.08em
		T\kern-.1667em\lower.7ex\hbox{E}\kern-.125emX}}

\begin{document}
	
	\title{Meta-UAD: A Meta-Learning Scheme for User-level Network Traffic Anomaly Detection}
	
	\author{
		\IEEEauthorblockN{Tongtong Feng$^1$, Qi Qi$^2$, Lingqi Guo$^2$, Jingyu Wang$^{2\star}$\thanks{$^{\star}$Corresponding author.}}
		\IEEEauthorblockA{$^1$ Tsinghua University, Beijing, China}
		\IEEEauthorblockA{$^2$ SKL-NST, Beijing University of Posts and Telecommunications, Beijing, China}
		\IEEEauthorblockA{fengtongtong@tsinghua.edu.cn, \{qiqi8266, guolingqi, wangjingyu\}@bupt.edu.cn}
	}
	
	\maketitle
	
	\begin{abstract}
		Accuracy anomaly detection in user-level network traffic is crucial for network security. Compared with existing models that passively detect specific anomaly classes with large labeled training samples, user-level network traffic contains sizeable new anomaly classes with few labeled samples and has an imbalance, self-similar, and data-hungry nature. Motivation on those limitations, in this paper, we propose \textit{Meta-UAD}, a Meta-learning scheme for User-level network traffic Anomaly Detection. Meta-UAD uses the CICFlowMeter to extract 81 flow-level statistical features and remove some invalid ones using cumulative importance ranking. Meta-UAD adopts a meta-learning training structure and learns from the collection of K-way-M-shot classification tasks, which can use a pre-trained model to adapt any new class with few samples by few iteration steps. We evaluate our scheme on two public datasets. Compared with existing models, the results further demonstrate the superiority of Meta-UAD with 15{\%} - 43{\%} gains in F1-score.
	\end{abstract}
	
	
	\section{Introduction}
	Accuracy anomaly detection in user-level traffic is crucial for network security\cite{1}. Attackers might exploit an application containing a vulnerability, jeopardizing the confidentiality, integrity, and availability of the user's crucial information. 
	
	User-level network traffic contains sizeable new anomaly classes with few labeled samples. Those new anomaly classes possess three unique characteristics. {\it Imbalanced}\cite{o8, 9472814}: compared to the existing anomaly classes, it is expensive and arduous to collect a massive amount of data onto the new anomaly class. Therefore, the sample sizes of different anomaly classes in the sampled database are often highly imbalanced. {\it Self-similar}\cite{o11, MTAD}: the new anomaly classes evolve from the existing ones and have characteristics closer to normal traffic. As our detection models adapt, so does anomaly traffic. According to the McAfee labs threats reports\footnote{https://media.mcafeeassets.com/content/dam/npcld/ecommerce/en-us/docs/reports/rp-mobile-threat-report-feb-2023.pdf.}, most new anomaly classes are branches of existing anomaly families. {\it Data-hungry}\cite{o13, U2UData}: now more attackers are focusing on fewer but more precise targets instead of widespread invasions, so each new anomaly class has small-scale samples. According to the CrowdStrike report\footnote{https://www.crowdstrike.com/global-threat-report/.}, near 62{\%} attackers will remain silent until a precise target is discovered, and they use legitimate credentials and built-in tools to attack, which can avoid detection by traditional detection models.
	
	\begin{figure}[t]
		\centering
		\includegraphics[width=0.95\linewidth]{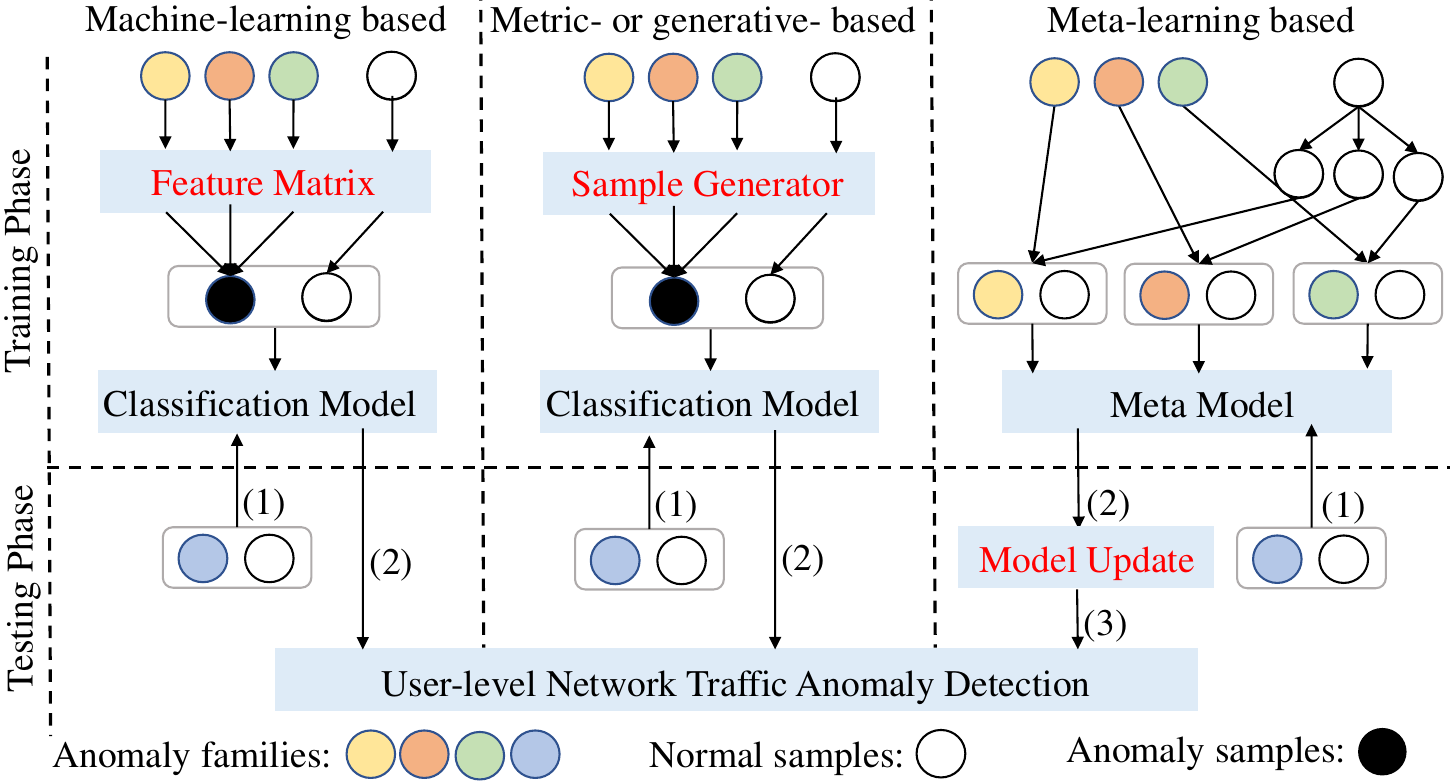}
		\caption{Network traffic anomaly detection schemes.}
		\label{FSL}
	\end{figure}
	
	Existing anomaly detection models can be grouped into three categories (see in Fig.\ref{FSL}). Machine-learning (ML) based models\cite{o1, o2, o3} classify data using a feature matrix where rows and columns correspond to each packet and feature values. They can combine the rules that differentiate anomaly traffic from normal traffic and achieve high-precision classification. Metric-based models\cite{TADAM, Simpleshot, o6} can learn a non-linear metric to generate new anomaly samples by distance to a class prototype or a class sub-space. Simultaneously, the generative-based models\cite{Delta-encoder, Mfcgan, o7} can learn a sample generator only from seen class samples to synthesize new samples using Generative Adversarial Networks (GAN).
	
	Existing models only detect specific anomaly classes with large labeled training samples, which are hard to detect anomaly samples in user-level network traffic. 1) They heavily rely on large-scale training samples and anomaly class distribution in training datasets\cite{o4}. They only can achieve good results for specific anomaly classes (same as the training dataset). For small-scale training datasets (e.g., endpoint traffic), the rules learned by these models may be entirely inapplicable to other datasets (e.g., cloud traffic). 2) For extremely imbalanced training datasets, models can produce an inductive learning bias towards the majority classes due to overfitting\cite{o9}, resulting in poor minority class detection performance. The sample generator can alleviate the data imbalance problem. Nevertheless, the sample generator uses only seen class data and does not explicitly learn to generate unseen class samples. If we detect many unseen classes, the number of synthesizing samples would be unfeasibly estimated\cite{o5}, and this operation will drastically increase computational complexity and energy consumption. In the testing phase, when the new anomaly class is significantly different from anomaly classes in the training set, the performance of those models is still poor\cite{o10}. 3) Since the anomaly classes are constantly updated and training data cannot cover all anomaly classes, the effectiveness of existing models will decline sharply for new anomaly classes\cite{o12}. 
	
	We argue that the recent breakthroughs in meta-learning have greatly facilitated small-scale sample processing capacities. Among them, few-shot learning\cite{FSL1} enables the model to effectively learn information from small samples by combining limited supervised information (few labeled samples) with prior knowledge (meta-network). Meta-SGD\cite{Meta-SGD} is a dominant optimization-based few-shot learning model. Motivation on this opportunity, in this paper, we propose \textit{Meta-UAD}, a Meta-learning scheme for User-level network traffic Anomaly Detection, which is based on Meta-SGD. Meta-UAD uses the CICFlowMeter\cite{CICFlowMeter} to extract 81 flow-level statistical features and remove some invalid ones using cumulative importance ranking. Meta-UAD adopts a meta-learning training structure and learns from the collection of K-way-M-shot classification tasks (see in Fig.\ref{FSL}). To evaluate our scheme's efficiency, we conduct two comparative experiments in two public network traffic datasets, and the results further demonstrate the superiority of Meta-UAD with 15{\%} - 43{\%} gains in the F1-score.

	\section{Anomaly Detection with Meta-learning}
	\subsection{Problem Definition}
	We define flow $f = \{p_1, p_2,...,p_u\}$ as a set of adjacent packets $p$ sharing the same 5-tuple information: [source IP address, destination IP address, source port number, destination port number, and transport layer protocol], where $u$ is equal to the number of the packet $p$ in flow $f$. We define $x_f = \{a_f^j\}_{0\leq j \leq v}$ as the feature vector of flow $f$, where $a_f^j$ is one feature of flow $f$, such as the packet size, and $v$ is the feature number of flow $f$. We denote by $y_f$ the detection label of flow $f$. Therefore, each sample in the training dataset can be expressed as ($x_f, y_f$), and each anomaly detection model can be shown: $\pi_{\theta}:x \rightarrow y$.
	
	In the user-level network traffic scenarios, we have a large training dataset $\mathcal{D}^{tr}$ (typically anomaly classes with large samples) and a test dataset $\mathcal{D}^{test}$ (new anomaly classes with one or a few samples), in which their respective class setscolon $\mathcal{C}_{tr}=\{1,...,|\mathcal{C}_{tr}|\}$ and $\mathcal{C}_{test}=\{|\mathcal{C}_{tr}|+1,...,|\mathcal{C}_{tr}|+|\mathcal{C}_{test}|\}$ are disjoint. Our problem is how to learn an anomaly detection model $\pi_{\theta}$ on $\mathcal{C}_{tr}$ that can accurately generalize to new anomaly classes $\mathcal{C}_{test}$ with few labeled samples by few computational consumptions.
	
	\subsection{Scheme Design}
	In this subsection, we describe Meta-UAD's design details, which is based on Meta-SGD\cite{Meta-SGD} to identify anomaly traffic.
	
	{\bf Feature extractor.} We first use CICFlowMeter\cite{CICFlowMeter} to extract 81 flow-level statistical features $\vec{A}_f^v$. Then we remove some invalid features: features' missing proportion is greater than 50{\%}, features' entropy is equal to 0, or features' importance ranking in RF\cite{RF} is in the bottom 30{\%} (all parameters are obtained through experimental testing). Finally, we select 33 flow-level features, as shown in TABLE \ref{feature set}.
	
	{\bf Task Definition.} We first sample $K$ classes from $\mathcal{C}_{tr}$, $\mathcal{C}^K \sim \mathcal{C}_{tr}$. In each episode of meta-training phase, we construct task $\mathcal{T}_i = \{\mathcal{D}^{sup}_i,\mathcal{D}^{val}_i\}$ for each class $\mathcal{C}^i$, where $\mathcal{C}^i \in \mathcal{C}^K$. For each episode, we build $K$ tasks. $\mathcal{D}^{sup}_i=\{(x_i,y_i)_m\}$ contain $M$ labelled flows from a anomaly class $\mathcal{C}^i$ and normal classes; $\mathcal{D}^{val}_i=\{(x_i,y_i)_n\}$ contain $N$ labelled flows from anomaly class $\mathcal{C}^i$ and normal classes.

	\begin{table}[t]
		\centering
		\fontsize{8pt}{12pt}\selectfont
		\caption{Extracting 33 flow-level statistical features}
		\label{feature set}
		\begin{threeparttable}
			\setlength{\tabcolsep}{4mm}{
				\begin{tabular}{|c|c|}
					\hline
					Feature ID & Feature Name  \\ \hline
					1-2 & Bwd packet count: Total, Mean  \\ \hline
					3-4  & Flow packet count: Total, Mean \\ \hline
					5 & Bwd/Fwd packet total count ratio \\ \hline
					6 & Fwd header Length: Total \\ \hline
					7 & Bwd header Length: Total \\ \hline
					8 & Flow header Length: Total \\ \hline
					9 & Bwd/Fwd header total length ratio \\ \hline
					10-13 & Fwd packet Length: Total, Max, Mean, Std  \\ \hline
					14-17 & Bwd packet Length: Total, Max, Mean, Std \\ \hline
					18-21 & Flow packet Length: Total, Max, Mean, Std \\ \hline
					22 & Bwd/Fwd packet total length ratio \\ \hline
					23-26 & Fwd IAT: Min, Max, Mean, Std \\ \hline
					27-28 & Bwd IAT: Max, Mean \\ \hline
					29-31 & Flow IAT: Total, Max, Mean\\ \hline
					32 & Fwd flag count: PSH \\ \hline
					33 & Flow flag count: ACK \\ \hline
			\end{tabular}}
			\begin{tablenotes}
				\footnotesize
				\item[1] Fwd/Bwd: forward/backward flow; 
				\item[2] IAT: the sending time interval between two adjacent packets. 
				\item[3] Flag count: the number of packets with this flag.
			\end{tablenotes}
		\end{threeparttable}
	\end{table}
	{\bf Meta-learning.} The goal of meta-learning is to construct a pre-trained model $\pi_{(\theta,\alpha)}:x \rightarrow y$ with parameters $\theta,\alpha$. $\theta$ is a set of the model's initial parameters; $\alpha$ is a vector that decides both the model's update direction and length. The meta-learning phase includes an inner update for a special task to update the local parameters and an outer update for all tasks to update global parameters.
	
	The inner update adjusts the local parameters $\theta'_i$ for task $\mathcal{T}_i$ through a gradient descent algorithm. For a given anomaly class $\mathcal{C}^i$ and task $\mathcal{T}_i$, we first definite a loss function on the support set $\mathcal{D}^{sup}_i$:
	\begin{equation}
		\mathcal{L}_{\mathcal{T}_i}(\pi_{\theta};\mathcal{D}^{sup}_i) = \sum_{(x_i,y_i)\in \mathcal{D}^{sup}_i}L(\pi_{\theta}(x_i),y_i)
	\end{equation}
	where $\pi_{\theta}(x_i)$ is a probability vector that shows the probability of sample $x_i$ being detected as different classes; $y_i$ is a one-hot vector that shows the actual class of the sample $x_i$; $|\pi_{\theta}(x_i)|=|y_i|=K+2$, it means that there are $K+2$ classes. Among them, there are $K$ anomaly classes in training tasks, one normal class, and one new anomaly class in testing tasks. We define $L(\cdot)$ as the cross entropy error function:
	\begin{equation}
		L(\pi_{\theta}(x_i),y_i) = -\sum_{k=1}^{K+2}y_i^klog(\pi_{\theta}(x_i)^k)
	\end{equation}
	where $\pi_{\theta}(x_i)^k$ is the probability that sample $x_i$ is predicted to the class $k$. Then, we use a gradient update algorithm to change the local parameters from $\theta$ to $\theta'_i$: 
	\begin{equation}
		\theta'_i=\theta - \alpha \circ \bigtriangledown_{\theta}\mathcal{L}_{\mathcal{T}_i}(\pi_{\theta};\mathcal{D}^{sup}_i)
		\label{E1}
	\end{equation}
	where the direction of $\alpha \circ \bigtriangledown_{\theta}\mathcal{L}_{\mathcal{T}_i}(\pi_{\theta};\mathcal{D}^{sup}_i)$ represents the direction of gradient update, and the length of $\alpha \circ \bigtriangledown_{\theta}\mathcal{L}_{\mathcal{T}_i}(\pi_{\theta};\mathcal{D}^{sup}_i)$ represents the length of gradient update.
	
	The outer update is to update the global parameters $\theta,\alpha$ by taking into account all the sampled tasks. For a given anomaly class $\mathcal{C}^i$ and task $\mathcal{T}_i$, we also definite a loss function on the validation set $\mathcal{D}^{val}_i$:
	\begin{equation}
		\mathcal{L}_{\mathcal{T}_i}(\pi_{\theta'_i};\mathcal{D}^{val}_i) = \sum_{(x_i,y_i)\in \mathcal{D}^{val}_i}L(\pi_{\theta'_i}(x_i),y_i)
		\label{E2}
	\end{equation}
	
	Then, we use a gradient update algorithm to change the parameters from $(\theta'_i, \alpha)$ to $(\theta',\alpha')$ by taking into account all the sampled tasks:
	\begin{equation}
		{(\theta', \alpha')} ={(\theta, \alpha)}-\beta\bigtriangledown_{(\theta, \alpha)}\sum_{i=1}^K\mathcal{L}_{\mathcal{T}_i}(\pi_{\theta'_i};\mathcal{D}^{val}_i)
	\end{equation}
	where $\beta$ is the learning rate of the outer update.

	{\bf Meta-testing.} During meta-testing, for a new anomaly class $\mathcal{C}^{new} \sim \mathcal{C}_{test}$, we first construct a support set $\mathcal{D}^{sup}_{new}$ and obtain the adapted parameters $\theta'$ with few iteration steps by the gradient update algorithm:
	\begin{equation}
		{\theta'} ={\theta}-\alpha \circ \bigtriangledown_{\theta}\mathcal{L}(\pi_{\theta}(\cdot);\mathcal{D}^{sup}_{new})
	\end{equation}
	
	Then, we use this updated model $\pi_{\theta'}(\cdot)$ for anomaly detection in new anomaly class $\mathcal{C}^{new}$. 
	
	{\bf Backbone architecture.} Our scheme Meta-UAD is general. In theory, we can use any anomaly detection network as the backbone architecture. Because the data format of network traffic is irregular, the features of network traffic have been extracted preliminarily through the data preprocessing module. For the anomaly detection module, we only need a simple backbone architecture to learn good accuracy by using our scheme. So we come up with the Deep Neural Networks (DNN) as backbone architecture.
	
	\begin{figure}[t]
		\centering
		\includegraphics[width=0.95\linewidth]{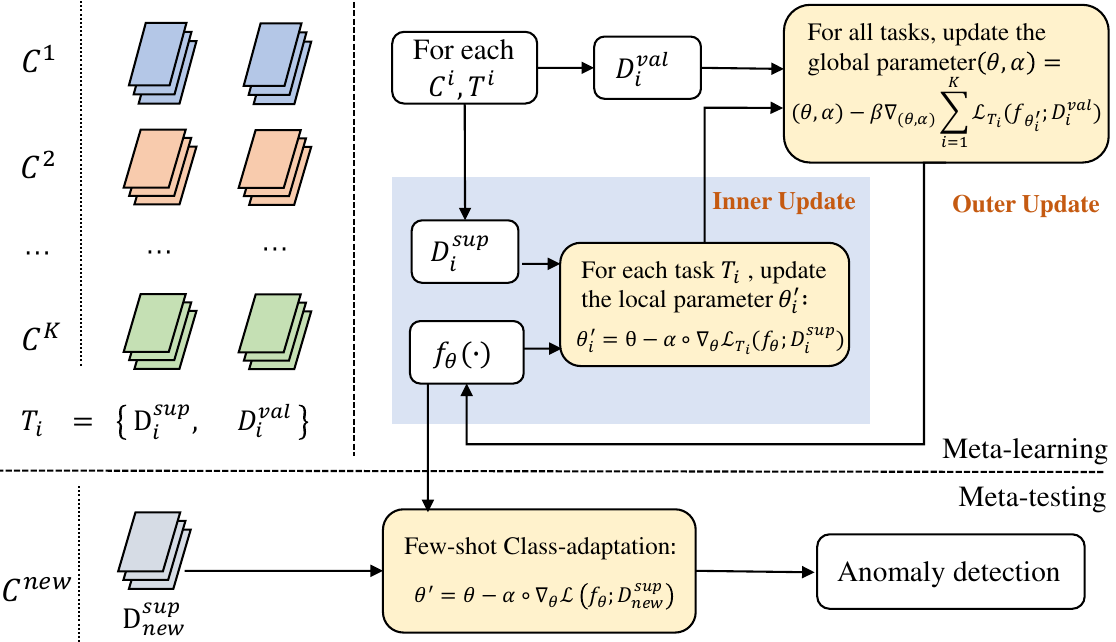}
		\caption{User-level network traffic anomaly detection with meta-learning.}
		\label{structure}
	\end{figure}
	
	\section{Experiments}
	\subsection{Experimental Setup}
	Experiments are conducted based on two public datasets: CIC-AndMal2017\cite{CICAndMal2017} and CIC-IDS2017\cite{CICIDS2017}. The CIC-AndMal2017 contains 4 main malware categories and 42 malware families with 1205515 normal flows and 1411064 anomaly flows. This dataset can be viewed as imbalanced. The imbalanced ratios of normal flows to Pletor, FakeJobOffer, AndroidDefender, and Gooligan are 255:1, 39:1, 21:1, and 13:1, respectively.  CIC-IDS2017 is provided by the Canadian Institute of Cybersecurity, which captures 2271320 normal flows and 556556 anomaly flows. Compared to CIC-AndMal2017, it is an extremely imbalanced dataset and it contains many novel attacks. The imbalanced ratios of normal flows to HeartBleed, XSS, and DoS are 206483:1, 3483:1, and 9:1, respectively.
	
	We use DNN as the backbone architecture for our scheme, which contains three fully connected layers with hidden cells $[256,128,128]$. Outputs from these layers are then aggregated in a hidden layer that uses $K+2$ neurons to apply the loss function. We fix the hyperparameters $\beta$ in meta-learning at $0.001$. During meta-training, we sample $K$ anomaly classes from training datasets, where $K=5$; we sample $M$ and $N$ flows per class, where $M=N$.
	
	Considering the data imbalanced condition in a multiclass classification problem, to efficiently judge our scheme's performance in detecting new anomaly classes with few labeled samples, we use F1-score as the criterion. We use five ML-based models as comparison algorithms: SVM \cite{SVM}, DT \cite{DT}, RF \cite{RF}, CNN \cite{CNN}, LSTM \cite{LSTM}. Similarly, we use two metric-based models and two generative-based models as comparison algorithms: Simpleshot\cite{Simpleshot}, TADAM\cite{TADAM}, Delta-encoder\cite{Delta-encoder}, and Mfc-gan\cite{Mfcgan}. 
	
	Suppose we have a training dataset $\mathcal{C}_{tr}$ and a test dataset $\mathcal{C}_{test}$. During the training phase, Meta-UAD samples $M$ labeled flows per class $\mathcal{C}^i \sim \mathcal{C}_{tr}$ in each episode to construct meta-models through inner update iteration and outer update iteration. Meta-UAD then uses a pre-training meta-model to adapt to new anomaly classes $\mathcal{C}^{new}$ with $M$ labeled samples through few iterative steps. Existing models use the entire training dataset and $M$ labeled samples of new anomaly classes $\mathcal{C}^{new}$ for model optimization until we get the optimal model on the considered evaluation metrics. 
	During the testing phase, we resample $M$ labeled flows of class $\mathcal{C}^{new} \sim \mathcal{C}_{test}$ for comparing between Meta-UAD and existing models. Since the sample number in the testing set is minimal, we choose to take the mean value of multiple experiments (100 times) to achieve the reliability of the results. 
	
	Moreover, we also define an additional baseline. {\it Pre-training:} the model learning from the meta-learning phase is directly applied to the meta-test phase without any adaptation. {\it Fine-tuning:} the model learning from the meta-learning phase is applied to the meta-test phase by a few iteration steps.
	
	\subsection{Meta-UAD on M-shot anomaly detection} 
	In this subsection, to verify that Meta-UAD can accurately detect new anomaly classes with few labeled samples, we compare Meta-UAD with the existing models on the M-shot network traffic anomaly detection in terms of F1-score. In the CIC-AndMal2017 dataset, we randomly select 30 classes from 42 anomaly families as the training set $\mathcal{C}_{tr}$ and the remaining classes as the testing set $\mathcal{C}_{test}$. We set $M = 5$, $M = 10$, $M = 20$, respectively. The experimental results are shown in TABLE \uppercase\expandafter{\romannumeral2}.
	
	Our scheme is far superior to existing models in terms of F1-score. ML-based models need to be trained in large training samples, to realize high-precision classification by summarizing the rules that distinguish anomaly traffic from normal traffic. They heavily depend on the dataset's distribution and only can achieve good results for specific anomaly classes with large samples. Therefore, those models will perform extremely poorly for a new anomaly class with few samples. Since the anomaly class in the testing set has few labeled samples, those models can produce an inductive learning bias towards the majority classes. Metric-based and generative-based models can learn a sample generator using only seen class data and do not explicitly learn to generate the new class samples. If we detect many new classes, the number of synthesizing samples would be unfeasibly estimated. In the testing phase, when the new anomaly class is significantly different from anomaly classes in the training set, the performance of those models is still poor and the computational cost is very high.
	
	We can find that the fine-tuning model is better than the pre-trained model. Before model testing, the fine-tuning model needs few iteration steps to adjust parameters and adapt to the specific new anomaly class. This operation is the core advantage of Meta-UAD.
	
	\subsection{Generalization} 
\begin{figure}[t]
		\centering 
		\subfigure[CIC-IDS2017]{
			\centering
			\includegraphics[width=0.49\linewidth]{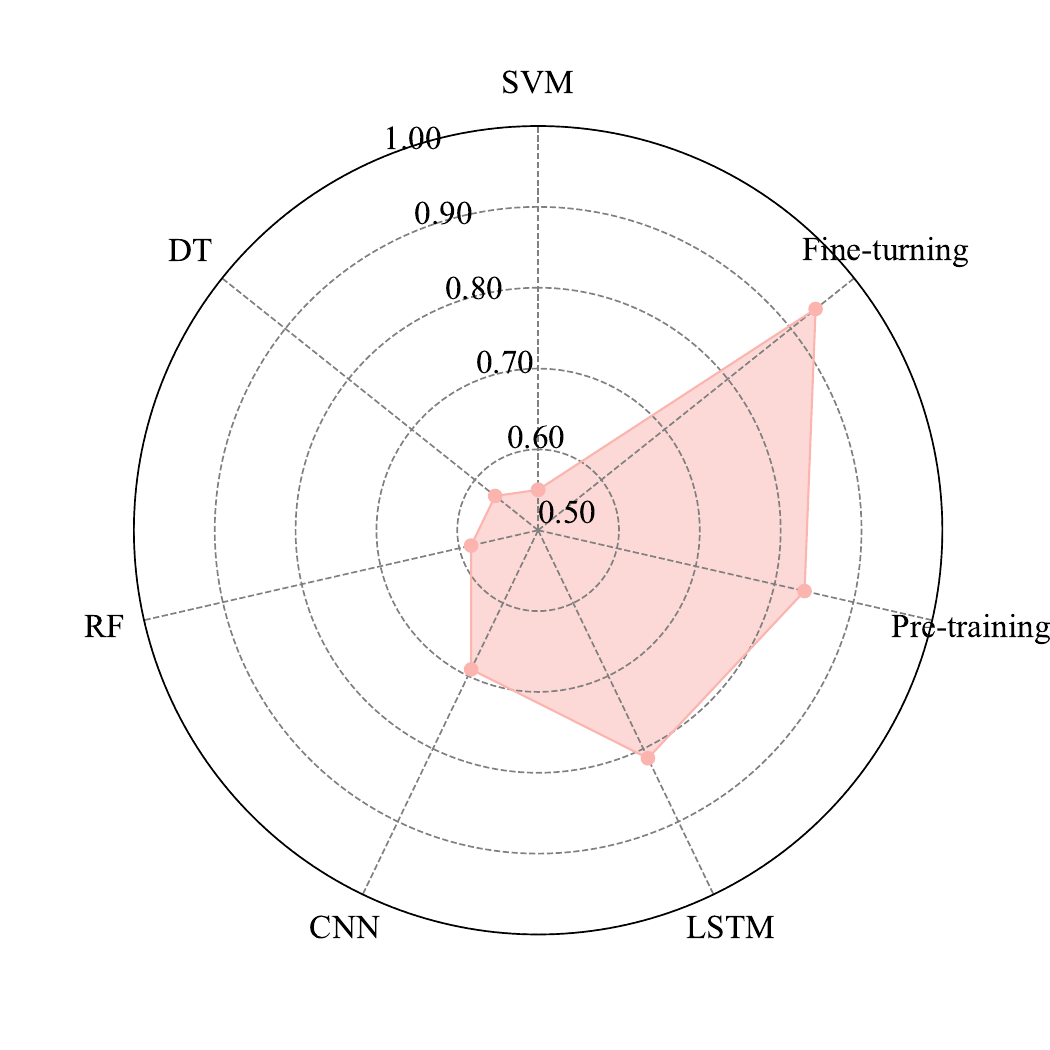}
			\label{plot_3_1}
		}%
		\subfigure[CIC-AndMal2017]{
			\centering
			\includegraphics[width=0.49\linewidth]{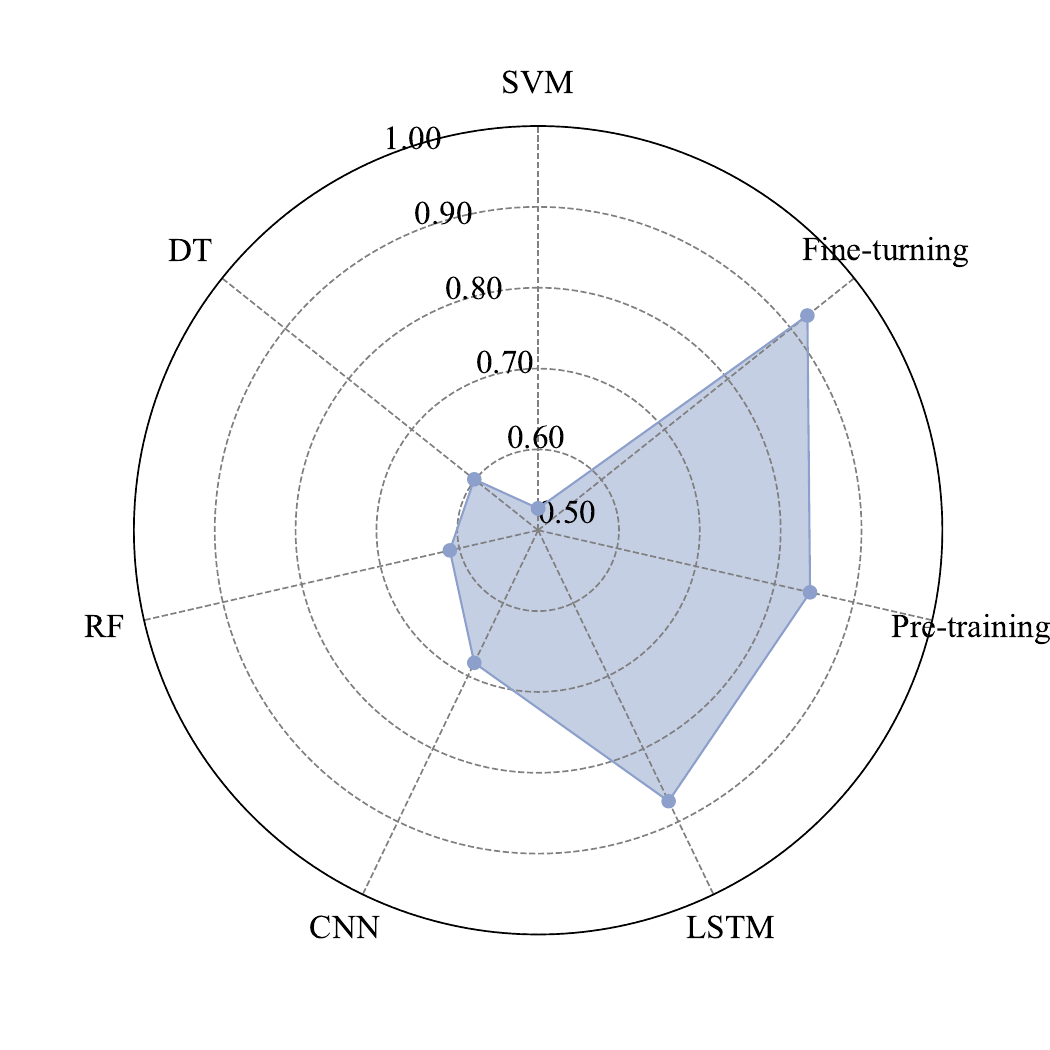}
			\label{plot_3_2}
		}%
		\caption{Compare our scheme with existing models in cross-dataset in terms of F1-score.}
	\end{figure}

	In this subsection, to verify that Meta-UAD has a strong generalization, we compared it with existing models in a cross-dataset in terms of F1-score. We select all anomaly families in CIC-AndMal2017 as the training set and anomaly families in CIC-IDS2017 as the testing set. We set $M=20$, and the experimental results are shown in Fig. \ref{plot_3_1}. When CIC-IDS2017 is the training set and CIC-AndMal2017 is the testing set, the experimental results are shown in Fig. \ref{plot_3_2}.
	
	We find that our scheme can still achieve optimal results in cross-dataset. Existing models heavily depend on the dataset's distribution and have poor generalization abilities to a new dataset. Our scheme needs to iterate some steps according to new anomaly samples in the meta-test phase, which can maximize the model's generalization and degrade the model's computational cost.
	
	\begin{table}[t]
		\centering
		\label{tab:my-table}
		\fontsize{8pt}{12pt}\selectfont
		\caption{Compare Meta-UAD with existing anomaly detection models on the M-shot anomaly detection task.}
		\begin{tabular}{c|c|ccc}
			\toprule
			\multirow{2}{*}{Category}         & \multirow{2}{*}{Algorithm} & \multicolumn{3}{c}{F1-score}            \\ \cline{3-5} 
			&                            & M=5            & M=10           & M=20  \\ \midrule
			\multirow{5}{*}{ML-based}         & SVM\cite{SVM}                        & 0.554          & 0.602          & 0.695 \\
			& DT\cite{DT}                         & 0.638          & 0.656          & 0.749 \\
			& RF\cite{RF}                        & 0.681          & 0.763          & 0.815 \\
			& CNN\cite{CNN}                        & 0.697          & 0.774          & 0.836 \\
			& LSTM\cite{LSTM}                       & 0.782          & 0.851          & 0.937 \\ \midrule
			\multirow{2}{*}{Metric-based}     & TADAM\cite{TADAM}                      & 0.834          & 0.868          & 0.895 \\
			& Simpleshot\cite{Simpleshot}                 & 0.848          & 0.879          & 0.934 \\ \midrule
			\multirow{2}{*}{Generation-based} & Delta-encoder\cite{Delta-encoder}               & 0.877          & 0.926          & 0.950 \\
			& Mfc-gan\cite{Mfcgan}                    & 0.831          & 0.894          & 0.929 \\ \midrule
			\multirow{2}{*}{Ours}                             & Pre-training               & 0.893          & 0.916          & 0.943 \\
			& \textbf{Fine-turning}               & \textbf{0.949} & \textbf{0.966} & \textbf{0.981} \\ \bottomrule
		\end{tabular}
	\end{table}
	
	\section{CONCLUSION}
	Anomaly detection aims to automatically identify anomaly behaviors by learning exclusively from normal traffic. Accuracy anomaly detection in user-level network traffic is crucial for network security. User-level network traffic contains sizeable new anomaly classes with few labeled samples and have an imbalance, self-similar, and data-hungry nature. For those limitations, we proposed Meta-UAD. It can use the pre-trained model to adapt new anomaly classes with few samples and going through few iteration steps. By experiment evaluation, our scheme are better than others. In the future, we hope to deploy our scheme into a real gateway.
	
	\section{Acknowledgments}
	This work was supported in part by the National Natural Science Foundation of China under Grants (62171057, 62471055, U23B2001), China Postdoctoral Science Foundation under Grant No. 2024M751688, Postdoctoral Fellowship Program of CPSF under Grant No. GZC20240827, and Beijing Key Lab of Networked Multimedia.
	
	\bibliographystyle{./IEEEtran}
	\balance
	\bibliography{IEEEabrv, Meta-UAD}
\end{document}